\documentclass[pageno]{jpaper}

\usepackage{mathptmx} 
\usepackage{fancyhdr}
\usepackage[normalem]{ulem}
\usepackage[keeplastbox]{flushend}
\usepackage[english]{babel}
\usepackage[utf8]{inputenc}
\usepackage{algorithm}
\usepackage{float}
\usepackage{wrapfig}
\usepackage[noend]{algpseudocode}
\usepackage{textcomp}
\usepackage[dvipsnames]{xcolor}
\usepackage{xspace}
\usepackage{subfig}
\usepackage{booktabs}
\usepackage{enumitem}
\usepackage[noadjust]{cite}
\usepackage{graphicx}
\usepackage{pifont}
\usepackage{setspace}
\usepackage{mwe}
\usepackage{multirow}
\usepackage[compact]{titlesec}
\usepackage{amsmath}
\usepackage{tikz}

\newcommand{\titleShort}{Mensa\xspace}
\newcommand{\exampleDesign}{\titleShort-G\xspace}
\newcommand{\accelA}{Pascal\xspace}
\newcommand{\accelB}{Pavlov\xspace}
\newcommand{\accelC}{Jacquard\xspace}

\makeatletter

\renewcommand\newblock{\hskip .11em\@plus.33em\@minus.07em}
\makeatother

\begin{document}

\bstctlcite{IEEEexample:BSTcontrol}

\title{\scalebox{0.98}{Heterogeneous Data-Centric Architectures for Modern Data-Intensive Applications:} \\
\scalebox{0.98}{Case Studies in Machine Learning and Databases}}

\newcommand{\affilETH}{$^\diamond$}
\newcommand{\affilGoogle}{$^\ddag$}
\newcommand{\affilUIUC}{$^\odot$}
\newcommand{\affilSFU}{$^\star$}

\author{
        \vspace{-15pt}\\ \scalebox{0.99}{Geraldo F. Oliveira}\affilETH~\quad%
        \scalebox{1.0}{Amirali Boroumand}\affilGoogle~\quad%
        \scalebox{1.0}{Saugata Ghose}\affilUIUC~\quad%
        \scalebox{1.0}{Juan Gómez-Luna}\affilETH~\quad%
        \scalebox{1.0}{Onur Mutlu}\affilETH%
    \\%
   \vspace{-10pt}\\% 
        \it\normalsize \affilETH ETH Z{\"u}rich  \qquad
        \affilGoogle Google \qquad
        \affilUIUC University of Illinois Urbana-Champaign
    \vspace{-10pt}%
}

%%%%%%%%%%%%%%%%%%%%%%%%%%%%%%%%%%%%
\date{}
\maketitle

\section{Motivation \& Problem}

Today's computing systems require moving data {back-and-forth between computing resources (e.g., CPUs, GPUs, accelerators) and off-chip} main memory so that computation can take place on the data. Unfortunately, this \emph{data movement} is a major bottleneck for system performance and energy consumption~\cite{google-pim,mutlu2013memory}. One promising {execution paradigm that alleviates} the data movement bottleneck in modern and emerging applications is {\emph{processing-in-memory}} (PIM)~\cite{damov,oliveira2021pimbench,pim-book,mutlu2020modern,ghose.ibmjrd19,mutlu2019enabling, mutlu2015research,mutlu2013memory,mutlu2021intelligent,mutlu2020intelligent,mutlu2015main}, where the cost of data movement to/from main memory is reduced by placing computation {capabilities} close to memory. In {the \emph{data-centric} PIM paradigm}, the  logic close to memory has access to data with significantly higher memory bandwidth, lower latency, and lower energy consumption than {processors/accelerators in existing \emph{processor-centric} systems.}

Naively employing PIM to accelerate data-intensive workloads can lead to sub-optimal performance due to the many design constraints PIM substrates impose (e.g.,  limited area and power budget available inside 3D-stacked memories~\cite{mutlu2020modern} or manufacturing limitations of combining memory and logic elements~\cite{mutlu2020modern,devaux2019true}). Therefore, many recent works \emph{co-design} specialized PIM accelerators and algorithms to improve performance and reduce the energy consumption of (i) applications from various application domains, such as graph processing~\cite{pim-enabled, tesseract, graphpim, song2018graphr, LazyPIM, conda, graphp, angizi2019graphs, angizi2019graphide, zhuo2019graphq,dai2018graphh,huang2020heterogeneous,besta2021sisa,nai2015instruction,dai2019graphsar,li.dac16,han2018novel,zhuo2021distributed,azarkhish2016design,xie2021spacea,imani2019digitalpim,han2017novel,zhou2019gram,zheng2020spara,liu2020regra,li2018graphia,kim2020things}, machine learning~\cite{tetris,neurocube,shafiee2016isaac,chi2016prime, google-pim,boroumand2021mitigating,amiraliphd,lee2022isscc,kwon202125,lee2021hardware,ke2021near,niu2022184qps,he2020sparse,peemen2013memory,angizi2018imce,deng.dac2018,eckert2018neural,imani2019floatpim,cho2020mcdram,shin2018mcdram,hajinazarsimdram,recnmp,park2021trim,wang2021rerec,wilkening2021recssd,huang2022practical,yuan2021forms,huang2022accelerating,imani2020dual,resch2020mouse,kwon2019tensordimm,kim2019nand,cao2020performance,nori2021reduct,das2018towards,long2018reram,rakin2018pim,long2020q}, bioinformatics~\cite{kim.bmc18,kim.psb18,kim2017genome,kim2016genome,NIM, cali2020genasm,ghiasi2022genstore,cali2022segram,angizi2019dna,gupta2019rapid,li2021pim,angizi2020exploring,chowdhury2020dna,angizi2020pim,kaplan2020bioseal,zhang2021pim,chen2020parc,singh2021fpga,alser2020accelerating,nag2019gencache,zhou2021ultra,wu2021sieve,xu2020aquoman}, {high-performance computing~\cite{singh2020nero,singh2021fpga,denzler2021casper,vermij2017boosting,syncron,fernandez2020natsa,singh2019napel,liu-spaa17,scheduling-gpu-pim,tom,gu.isca16,pim-graphics,wang2019reram}}, databases~\cite{mondrian, santos2017operand, ambit, LazyPIM, conda,PICA,boroumand2021polynesia,amiraliphd,li.dac16, seshadri2019dram,seshadri.bookchapter17.arxiv,seshadri.bookchapter17,seshadri.arxiv16,hajinazarsimdram,Mingyu:PACT,JAFAR,lu2019agile,tome2018hipe,kepe2019database,sun2018bidirectional,lee2020optimizing,DBLP:conf/sigmod/BabarinsaI15,lekshmi2021coprao}, {security~\cite{gu2016leveraging,kim2019d,kim2018dram,xiong2022secndp,nejatollahi2020cryptopim,reis2020computing,li2019leveraging,glova2019near,bostanci2022dr,olgun2021quac},} data manipulation~\cite{ambit,li.micro17,seshadri2013rowclone, wang2020figaro, chang.hpca16, li.dac16, rezaei2020nom, and-or-in-dram}, and mobile workloads~\cite{google-pim,amiraliphd}; and 
(ii) {various} {computing} environments, {including} cloud systems~\cite{awan2017identifying,mondrian,tesseract,gomez2021benchmarking,gomez2021benchmarkingcut,devaux2019true,gomez2022benchmarking}{, mobile systems~\cite{google-pim}, and} edge devices~\cite{simon2020blade,si2019circuit}.

{We} showcase the benefits {of} co-designing algorithms and hardware {in a way that efficiently takes advantage of the PIM paradigm} for two {modern data-intensive applications}: 
(1) {machine learning {inference} models for edge devices and} 
(2) {hybrid transactional/analytical processing databases for cloud systems}. {We follow a two-step approach in our system design. {In the first step}, we extensively analyze the computation {and memory access} patterns of each application to gain insights into {its} hardware/software requirements and {major} sources of performance and energy {bottlenecks in} processor-centric systems. {In the second step}, we {leverage the insights from the first step to co-design algorithms and} hardware {accelerators} to {enable} high-performance and energy-efficient data-centric architectures for each application.} 

\section{Mensa: Accelerating Google Neural Network Models for Edge Devices}

Modern consumer devices make widespread use of
machine learning (ML). The growing complexity of these devices, combined with increasing demand for privacy, connectivity, and real-time responses, has spurred significant interest in pushing ML inference computation to
the edge (i.e., in or near consumer devices, instead of the cloud)~\cite{edge-facebook, edge-nature,eyerissv2}.
Due to their resource-constrained nature, edge computing platforms now employ specialized
energy-efficient accelerators for on-device inference
(e.g., Google Edge Tensor Processing Unit, TPU~\cite{edge-tpu}; NVIDIA Jetson~\cite{jetson};
Intel Movidius~\cite{movidius}). At the same time, neural network (NN) algorithms are evolving rapidly,
which has led to many types of NN models.

Despite the wide variety of NN model types, Google's state-of-the-art Edge TPU~\cite{edge-tpu} provides an optimized 
\emph{one-size-fits-all} design (i.e., a monolithic accelerator with a fixed, large  number of processing elements (PEs) and a fixed \emph{dataflow},
which determines how data moves within
the accelerator)
that caters to edge device area
and energy constraints. Unfortunately,  it is very challenging to achieve
high energy efficiency,
computational throughput, and
area efficiency for each NN model with this one-size-fits-all design. 

We conduct an in-depth analysis of ML inference execution on a commercial Edge TPU, across 24 state-of-the-art Google edge NN models spanning four popular NN model types:
(1)~convolutional neural networks (CNNs), 
(2)~long short-term memories (LSTMs)~\cite{lstm-google}, 
(3)~Transducers~\cite{he.icassp2019, transducer3, transducer4}, and 
(4)~recurrent CNNs (RCNNs)~\cite{rcnn-google,lrcn}. The \emph{key takeaway} from our extensive analysis of Google edge NN models on the Edge TPU is that \emph{all key components} 
of an edge accelerator (i.e., PE array, dataflow, memory system)
must be co-designed and co-customized based on specific layer characteristics to achieve high utilization and energy efficiency.
Therefore, our \emph{goal} is to revisit the design of edge ML accelerators such that they are aware of and can fully exploit
the growing variation within and across edge NN models.

To this end, we propose \titleShort, the first general HW/SW composable framework for ML acceleration in edge devices. The key idea of \titleShort is to perform NN layer
execution across \emph{multiple} on-chip and near-data accelerators,
each of which is small and tailored to the characteristics of a particular
subset (i.e., family) of layers. Our  experimental study of the characteristics of different layers in the Google edge NN models reveals that the layers naturally group into a small number of
clusters that are based on a subset of these characteristics.
This new insight allows us to significantly limit the number of different accelerators 
required in a \titleShort design. We design a runtime scheduler for \titleShort to determine which of these
accelerators should execute which NN layer,
using information about
(1)~which accelerator is
best suited to the layer's characteristics, and 
(2)~inter-layer dependencies.

Using our new insight about layer clustering, we develop \emph{\exampleDesign}, an example design for \titleShort optimized for Google edge NN models.
We find that the design of \exampleDesign's underlying accelerators should center around two
key layer characteristics (memory boundedness, and 
activation/parameter reuse opportunities). This allows us to provide efficient inference execution
for \emph{all} of the Google edge NN models using \emph{only three} accelerators in \exampleDesign {that we call \emph{\accelA}, \emph{\accelB}, and \emph{\accelC}.} \accelA, for compute-centric layers,
maintains the high PE utilization that these layers
achieve in the Edge TPU, but does so using an
optimized dataflow that both reduces the size of
the on-chip buffer (16$\times$ smaller than in the Edge TPU)
and the amount of on-chip network traffic.
\accelB, for LSTM-like data-centric layers,
employs a dataflow that enables the temporal reduction
of output activations, and
enables the parallel execution of layer operations in
a way that increases parameter reuse,
greatly reducing off-chip memory traffic.
\accelC, for other data-centric layers,
significantly reduces the size of the on-chip parameter buffer (by 32$\times$) using a dataflow that exposes reuse opportunities for parameters. As both \accelB and \accelC are optimized for data-centric layers,
which require significant memory bandwidth and are unable to utilize a significant fraction of PEs in the Edge TPU,
we place the accelerators in the logic layer of 3D-stacked
memories~\cite{hbm, hmcspec2, lee2016smla} and use significantly smaller PE arrays
compared to the PE array in \accelA, unleashing significant performance \emph{and} energy benefits.

Our evaluation shows that compared to the baseline Edge TPU, \exampleDesign reduces total inference energy by 66.0\%, improves energy efficiency (TFLOP/J) by 3.0$\times$, and increases computational throughput (TFLOP/s) by 3.1$\times$, averaged across all 24 Google edge NN models. \exampleDesign improves inference energy 
efficiency and throughput 
by 2.4$\times$ and 4.3$\times$ over Eyeriss~v2, a state-of-the-art accelerator.

{We conclude that employing our \titleShort framework and tailoring ML accelerators to the key characteristics of NN layers can provide performance, energy, and area benefits to edge devices.} {For more information on Mensa, a detailed description of  \exampleDesign's accelerators, and our extensive evaluation results, please refer to the full version of our paper~\cite{boroumand2021mitigating,boroumand2021google}.}

\section{Polynesia: Accelerating Hybrid Transactional/Analytical Processing using PIM}

Many application domains, such as fraud detection~\cite{cao2019titant,qiu2018real,quah2008real}, business intelligence~\cite{sql-htap,snappy-data,sahay2008real}, healthcare~\cite{chisholm2014adopting,ta2016big}, personalized recommendation~\cite{wiser,zhou2017kunpeng}, and IoT~\cite{wiser}, have a critical need to perform 
\emph{real-time data analysis}, where data analysis needs to be performed using the most recent version of data~\cite{real-time-analysis-sql,huang2020tidb}. To enable real-time data analysis, state-of-the-art database management systems (DBMSs) leverage \emph{hybrid transactional and analytical processing} (HTAP)~\cite{htap-gartner,sap-hana-evolution,htap}. An HTAP DBMS is a single-DBMS
solution that supports both transactional and analytical
workloads~\cite{htap-gartner,peloton,batchdb,htap-survey,real-time-analysis-sql}. 

Ideally, an HTAP system should have three properties~\cite{batchdb} to guarantee efficient execution of transactional and analytical workloads.
First, it should ensure that both transactional and analytical workloads benefit from their own workload-specific optimizations (e.g., algorithms, data structures).
Second, it should guarantee data freshness and data consistency (i.e., access to the most recent version of data) for analytical workloads
while ensuring that both transactional and analytical workloads have a consistent view of data across the system. Third, it should ensure that the latency and throughput of both the transactional workload and
the analytical workload are the same as if each of them were run in isolation.

We extensively study state-of-the-art HTAP systems and observe two key problems that prevent them from achieving all three
properties of an ideal HTAP system.
First, these systems experience a drastic reduction in
transactional throughput (up to 74.6\%) and analytical throughput (up to 49.8\%) compared to when transactional and analytical workloads run in isolation. This is because the mechanisms used to provide data freshness and consistency
induce a large amount of data movement between the CPU cores and
main memory. Second, HTAP systems often fail to provide effective performance isolation. These systems suffer from severe performance interference because of the high resource contention
between transactional workloads and analytical workloads. Therefore, \emph{our goal} is to develop an HTAP system that overcomes these problems while achieving all three of the
desired HTAP properties.

We propose a novel system 
for HTAP databases called 
\emph{Polynesia}. The key insight behind Polynesia is to partition the computing resources into two isolated new custom processing \emph{islands}: \emph{transactional islands} and \emph{analytical islands}. An island is a hardware--software co-designed component specialized for specific types of queries. Each island consists of (1)~a replica of data for a specific workload, 
(2)~an optimized execution engine, and
(3)~a set of hardware resources
that cater to the execution engine and its memory access patterns. 

\begin{figure}[ht]
    \centering
        \centering
        \includegraphics[width=\linewidth]{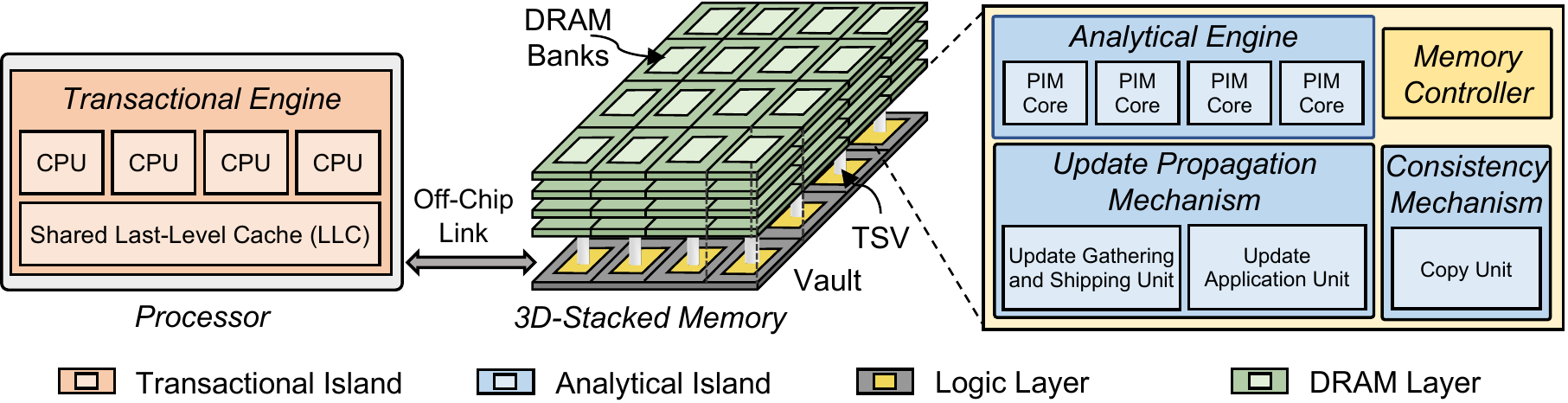}%
    \caption{High-level organization of Polynesia hardware.}
    \label{fig:high-level-hw}
\end{figure}

{Figure~\ref{fig:high-level-hw} shows the high-level organization of Polynesia, which includes one transactional island and one analytical island. Polynesia is equipped with
a 3D-stacked memory similar to the Hybrid Memory Cube (HMC)~\cite{hmcspec2},
where multiple vertically-stacked DRAM layers are connected with a 
\emph{logic layer} using thousands of \emph{through-silicon vias} (TSVs). 
An HMC chip is split up into multiple \emph{vaults}, where each vault corresponds to a vertical slice of the memory and logic layer.}

\pagebreak 
Polynesia meets all desired properties from an HTAP system in three ways. First, by employing processing islands, Polynesia enables workload-specific optimizations for both transactional and analytical workloads. Second, we design new hardware accelerators to add specialized capabilities to
each island, which we exploit to optimize the performance of several key HTAP algorithms. This includes new accelerators and modified algorithms to 
propagate transactional updates to analytical islands and to maintain a consistent view of data across the system. Such new components ensure data freshness and data consistency in our HTAP system.
Third, we tailor the design of transactional and analytical islands to fit the characteristics of transactional and analytical workloads. The transactional islands use dedicated CPU hardware resources (i.e., multicore CPUs and multi-level caches) to execute transactional workloads since transactional queries have cache-friendly access patterns~\cite{conda,LazyPIM,amiraliphd}. The analytical islands leverage PIM techniques~\cite{mutlu2020modern,pim-survey,ghose.ibmjrd19} due to the large data traffic analytical workloads produce. We equip the analytical islands with a new PIM-based analytical engine that includes simple in-order PIM cores added to the logic layer of a 3D-stacked memory~\cite{hmcspec2,kwon202125,lee2016smla}, software to handle data placement, and runtime task scheduling heuristics. Our new design enables the execution of transactional and analytical workloads at low latency and high throughput.

In our evaluations, we show the benefits of each component of Polynesia,
and compare its end-to-end performance and energy usage to three
state-of-the-art HTAP systems (modeled after Hyper~\cite{hyper}, AnkerDB~\cite{ankerdb}, and Batch-DB~\cite{batchdb}).
Polynesia outperforms all three, with higher 
transactional throughput (2.20$\times$/1.15$\times$/1.94$\times$; mean of 1.70$\times$) and 
analytical throughput (3.78$\times$/5.04$\times$/2.76$\times$; mean of 3.74$\times$), while consuming 48\% lower energy than the prior lowest-energy HTAP system. 

{We conclude that Polynesia efficiently provides high-throughput real-time data analysis, while meeting all three desired HTAP properties.} {For more information on Polynesia design, including our tailored algorithms and hardware units to maintain data freshness and data consistency, and the design of our analytical engine, {as well as our extensive evaluation results,} please refer to the full version of our paper~\cite{boroumand2022polynesia,boroumand2021polynesia}.} We open-source Polynesia and the complete source code of our evaluation~\cite{polynesia.github}.

\section{Conclusion \& Future Work}

{This paper summarizes our recent works on accelerating emerging data-intensive applications with processing-in-memory (PIM). We} showcase the performance and energy benefits that PIM provides for edge neural network models and hybrid transactional/analytical processing systems. Our proposed PIM architectures outperform state-of-the-art solutions by co-designing algorithms and hardware while reducing area costs. We hope our work inspires the design of novel, high-performance, and energy-efficient data-centric PIM architectures for many other emerging applications.

\section*{Acknowledgments}

We thank SAFARI Research Group members for valuable feedback and the stimulating intellectual environment they provide. {We acknowledge the generous gifts provided by our industrial partners, including ASML, Facebook, Google, Huawei, Intel, Microsoft, and VMware.
We acknowledge support from the Semiconductor Research Corporation and the 
ETH Future Computing Laboratory.} 

This invited extended abstract is a summary version of our prior works{, Mensa}~\cite{boroumand2021google,boroumand2021google_arxiv} (published at PACT 2021) and {Polynesia}~\cite{boroumand2022polynesia,boroumand2021polynesia} (published at ICDE 2022). Presentations {describing Mensa and Polynesia can be found at}~\cite{mensa_talk} and~\cite{polynesia_talk}{, respectively.}

%%%%%%%%% -- BIB STYLE AND FILE -- %%%%%%%%
\footnotesize
  \bibliographystyle{IEEEtran}
  \bibliography{refs}
%%%%%%%%%%%%%%%%%%%%%%%%%%%%%%%%%%%%

\end{document}